\documentstyle[12pt]{article}
\def\l{\noindent}
\def\0.5{\hspace{0.5cm}}
\def\1{\hspace{1cm}}
\def\2{\hspace{2cm}}
\def\v1{\vspace{1cm}}
\begin{document}
\baselineskip=20pt
\title{Stationary Axisymmetric Solutions of the Einstein Equations
 with Rigidly Rotating
Perfect Fluid and Nonlinear Charged Sources
\thanks{Supported in part by CONACyT.}}
\author{Humberto Salazar \\
FCFM, Benem\'erita Universidad Aut\'onoma de Puebla,\\
A.P. 1152, 72000 Puebla, M\'exico.
\and
Rub\'en  Cordero \\ 
Departamento de F\'{\i}sica, CINVESTAV-IPN,\\
A.P. 14-740, 07000 M\'exico, D.F., M\'exico.}
\date{}
\maketitle
\begin{abstract}
\noindent A class of stationary rigidly rotating perfect
fluid coupled with non-linear
electromagnetic fields was investigated. An exact solution of the
Einstein equations with sources for the Carter B(+) branch was found,
for the equation of state $3p + \epsilon = constant$.
We use a structural function for the Born-Infeld non-linear electrodynamics
which is invariant under duality rotations and a metric possessing a
four-
parameter group of motions. The solution is of Petrov type D and the
eigenvectors of the electromagnetic field are aligned to the
Debever-Penrose vectors.
\vspace{0.7cm}

\l PACS: 04.20.Jb; 04.40.+c
\end{abstract}
\newpage
\l {\sf I. BORN-INFELD NON-LINEAR ELECTRODYNAMICS
AND DUALITY ROTATIONS}

\v1
 The basic description of the dynamical equations of non-linear
electrodynamics within general relativity can be done in terms of the
null tetrad formalism according to which the metric is given by

\begin{equation}
 g=2e^{1}\otimes e^{2} + 2e^{3}\otimes e^{4},\hspace{1cm}
 e^{2}=\overline{e^{1}}
\end{equation}

\l where the $e^a\in \Lambda^{1}$ fulfill the Cartan structure equations
\begin{equation}
de^{a} = e^{b}\wedge \Gamma^{a}_{b}
=\Gamma^{a}_{bc}e^{b}\wedge e^{c}
\end{equation}
and $\Gamma^{a}_{b}\in \Lambda^{1}$ satisfy the second structure equations
\begin{equation}
d\Gamma^{a}_{b} +
\Gamma^{a}_{s}\wedge\Gamma^{s}_b=\frac{1}{2}R^{a}_{bcd}
e^{c}\wedge e^{d}
\end{equation}

 The Riemann curvature components $R^{a}_{bcd}$ may be replaced by the Weyl
conformal tensor components, which are characterized by five complex
curvature coefficients $C^{(a)}$, and the components of the traceless Ricci
tensor $ C_{ab}=R_{ab} - 1/4g_{ab}R$ where
$R_{ab}=R^{s}_{abs}$ and
$R=R^{a}_{a}$.
Non-linear theories of Born Infeld type (B-I) are theories with a
hamiltonian function ${\cal H} $ 
depending on the invariants of the skew-symmetric
tensor $P_{ab}$,

$$ P=\frac{1}{4} P^{ab}P_{ab}  \hspace{2cm}  Q=\frac{1}{4} \check{P}^{ab} P_{ab} $$

$$\check{P}^{ab}=-\frac{1}{2} \epsilon ^{abcd} P_{cd}  $$

\noindent where $\epsilon ^{abcd} $ is the Levi-Civita symbol with $\epsilon ^{1234} =1$. 

 Skew-symmetric tensor
$F_{ab}$ can be defined by the material equations
\begin{equation}
F_{ab} = {\cal H}_{P}P_{ab} + {\cal H}_{Q}\check{P}_{ab}
\end{equation}
\[
{\cal H}_{P} = \frac{\partial {\cal H}}{\partial P} \hspace{1cm} 
{\cal H}_{Q} = \frac{\partial {\cal H}}{\partial Q}
\]
We select the null tetrad in such a manner that out of all independent
components of the electromagnetic field tensors $F_{ab}$ (corresponding to
$\vec{E}$ and $\vec{B}$) and $P_{ab}$ (corresponding to $\vec{D}$ and
$\vec{H}$), they are different from zero only:

\begin{equation}
P_{34} = D,\hspace{0.5cm} P_{12} = iH,\hspace{0.5cm}F_{34} = E,
\hspace{0.5cm} F_{12} =iB
\end{equation}

\l where $D$, $H$, $E$, and $B$ are real.
The above selection for $F_{ab}$ and $P_{ab}$ can be made simultaneously
by virtue of the material equations.
The invariants of $F_{ab}$ and $P_{ab}$ read:

\begin{eqnarray}
\frac{1}{4} P^{ab}P_{ab} + \frac{1}{4} \check{P} ^{ab} P_{ab}=P + Q = -\frac{1}{2}(D + iH)^{2} \neq 0\\  \nonumber
\frac{1}{4} f^{ab}f_{ab} + \frac{1}{4} \check{f} ^{ab} f_{ab}=F + G = -\frac{1}{2}(E + iB)^{2} \neq 0
\end{eqnarray}

So that $(D,H)$ and $(E,B)$ can be interpreted as independent parameters of
the complex invariant of the electromagnetic field. We now can introduce
by a Legendre transformation of the Hamiltonian
${\cal H}$(P,Q) or ${\cal H}$(D,H) a new
structural function for the non-linear electrodynamics given by (Salazar et al., 1987)

\begin{equation}
M(D,B) = BH-{\cal H}(D,H)
\end{equation}
\noindent In the following we are going to use the same notation that Salazar et al. (1987)

\noindent Furthermore we restrict our structural function $M$ to be
invariant under duality rotations i.e.

\begin{equation}
M(D',B') = M(D,B) \hspace{1cm}{\rm for} \hspace{1cm}D'+iB' = e^{is}(D + iB)
\end{equation}

The last condition can be easily seen to constrain the function $M$ to be
a function of the variable $(D^{2} +B^{2})$ only:

\begin{equation}
M=b^{2}f(X),\hspace{1cm} X \equiv  \frac{1}{2b^{2}}(D^{2} +B^{2}),\hspace{1cm}
 b =constant.
\end{equation}

The original Born-Infeld theory given by the Hamiltonian function

\begin{equation}
{\cal H} = b^{2}-\sqrt{b^{4}- 2b^{2}P +Q^{2}}
\end{equation}
\noindent belongs to this class of theories invariant under duality rotations and
corresponds to

\begin{equation}
f(X)= \sqrt{1 + 2X} - 1
\end{equation}

\v1
\l {\sf II. EINSTEIN BORN INFELD EQUATIONS WITH PERFECT FLUIDS}

\v1
In this work we are concerned with solutions to the Einstein-Born-Infeld
equations with a perfect fluid

\begin{eqnarray} \nonumber
R_{ab}-\frac{1}{2}g_{ab}R &=&-T_{ab} \\ \nonumber
T_{ab}&=&(p+\epsilon )u_{a}u_{b}+pg_{ab}-8 \pi E_{ab} \\ \nonumber
u_{a}u^{a}&=&-1 \\
p + \epsilon &>& 0\\ \nonumber
4\pi E_{ab} &=&{\cal H}_{p}(-P_{as}P^{s}_{b} + g_{ab}P)
+ (P{\cal H}_{P} + Q{\cal H}_{Q}- {\cal H})g_{ab}\\
\nonumber
\check{F} _{ab;}\hspace{0.02cm} ^{a} &=& 0 \\ \nonumber
 P^{ba;}\hspace{0.02cm} _{a} &=& 4\pi J^{b}\nonumber
\end{eqnarray}

\noindent where $E_{ab}$ is the energy-momentum tensor of the non-linear
electromagnetic field, $u_{a}$ is the fluid four-velocity, $p$ is the
fluid pressure and $\epsilon$ is the energy density.
We shall consider the Carter type D metric with conformal factor and
having a four-parameter group of symmetries.
We shall work with the standard gravitational units so chosen
that the gravitational constant $G$ and the velocity of light $c$ are equal
to the unity. Let now $\varepsilon$ be a dimensionless constant, $l$ be a
constant of dimension of length, and ${x^{\mu}}:={\xi,\overline \xi,r,\tau}$
and ${x^{\mu}}:={u,v,r,\sigma}$ be two coordinate charts:  $\xi$ is
complex,
 while the remaining coordinates and constants are real. The $r$,
$\tau$ and $\sigma$ are of dimension of length while $\xi$, $u$ and $v$
are dimensionless.
Given these ingredients, we construct first the two 2-dimensional
Riemannian spaces of constant curvature

\[
\Lambda^{1}\otimes\Lambda^{1} \ni dl^{2(\pm)}:=\left\{
 \begin{array}{ll}
4\frac{d\xi\otimes d\overline \xi}{(1+\varepsilon \xi\overline
\xi)^{2}}\\
 4\frac{du\otimes dv}{(1+\varepsilon u v)^{2}}
\end{array}
\right. \]

\noindent given in terms of stereographic coordinates
 $(\xi,\overline \xi)$ and the
associated real one-forms of dimension of length

\[
\Lambda^{1}\ni\pi^{\pm}:=\left\{
\begin{array}{ll}
d\tau+2il(\frac{\overline \xi d\xi- \xi d\overline \xi}{1+
\varepsilon \xi\overline
\xi})\\
d\sigma + 2l(\frac{vdu-udv}{1+\varepsilon \xi\overline \xi})
\end{array}
\right. \]

 Then, introducing two real analytic functions $N = N(r)$ and
$F^{(\pm)}=F^{(\pm)}(r)$ (are of dimension (lenght)$^{2}$), we define now the 4-dimensional
Riemannian space-time of
signature $(+++-)$

\begin{equation}
B^{(\pm)}:ds^{2}:=Ndl^{2(\pm)}+\frac{N}{F^{(\pm)}}dr \otimes dr {\mp} \frac{F^{(\pm)}}{N}\pi^{(\pm)}\otimes \pi^{(\pm)}
\end{equation}

It can be easily shown that the (Carter, 1968) separable $B^{(\pm)}$ branches
of type D can be brought --- without any loss of generality --- to the form
of (13).
In the considered representation of the $B^{(\pm)}$ metrics, only the
sign of the parameter $\varepsilon$ is relevant.

\indent We should like now to explain why we consider (13) as the optimal
coordinatization of the Carter $B^{(\pm)}$ branches for our purpose. We
observe first that a formal transformation

\begin{equation}
z\rightarrow -u,\hspace{0.5cm} \overline z \rightarrow -v,
\hspace{0.5cm}
 \tau \rightarrow i\sigma,\hspace{0.5cm}
F^{(+)} \rightarrow F^{(-)}
\end{equation}

\l which obviously implies $dl^{2(+)}\rightarrow dl^{2(-)}$ ,
$\pi^{(+)}\rightarrow i\pi^{(-)}$ brings the $B^{(+)}$ metric into
the $B^{(-)}$ metric. We will see that this leads to a useful
computational advantage; the natural tetrads; connection and curvatures of
$B^{(\pm)}$ metrics can be treated via perfectly parallel computations.
Secondly, we notice that our coordinatization of the $B^{(\pm)}$ metrics
allows us to give a unified description of their minimal sub-group of
symmetries, which are 4-dimensional Lie Groups.
Next, taking the $B^{(+)}$ metrics, we should like to
investigate the advantages
of our coordinatization from the point of view of the
Debney-Kerr-Schild formalism described in the first section.
With the metrics (13) a natural choice for the null tetrads is
correspondingly

\[
     -\frac{\sqrt{2N}}{1+\varepsilon \xi\bar\xi} \left\{\begin{array}{l}
     d\xi \\
     d\bar\xi
     \end{array}
     \right.
     =\left\{ \begin{array}{c}
e^{1}\\
e^{2}
\end{array}
\right. \]

\[
 \frac{1}{\sqrt2}\left [ \left ( \sqrt{\frac{N}{F}} \right )dr \pm \sqrt{\frac{F}{N}} \pi \right]
 = \left\{ \begin{array}{c}
e^{3}\\
e^{4}
\end{array}
\right. \]

\noindent whereas the connection forms are given by

\begin{equation}
\Gamma_{42}=\sqrt{\frac{F}{2 N^3}}\left ( il-1/2\dot{N}\right ) e^1
\end{equation}

\begin{equation}
\Gamma_{31}=\sqrt{\frac{F}{2 N^3}}\left ( il-1/2\dot{N}\right ) e^2
\end{equation}
\noindent where dots denote the r derivative
\begin{equation}
\Gamma_{12} + \Gamma_{34} = \frac{\varepsilon}{\sqrt{2 N^3}}
\left [ \xi e^2-\bar\xi e^1 \right ]
-\frac{1}{\sqrt 2}\left [ \left ( \sqrt{\frac{F}{N}}\right ) ^{\dot {}}
  +  il\sqrt{\frac{F}{N^3}}\hspace {2mm} \right ]  \left ( e^3-e^4 \right )
\end{equation}

From the second structure equations
\begin{equation}
d\Gamma_{42} +\Gamma_{42}\wedge(\Gamma_{12}+\Gamma_{34})=
\gamma e^3\wedge e^1 + \delta e^4\wedge e^1
\end{equation}

 \begin{equation}
d\Gamma_{31}+(\Gamma_{12}+\Gamma_{34})\wedge \Gamma_{31}=
\delta e^3\wedge e^2 + \gamma e^4\wedge e^2
\end{equation}

\begin{equation}
d(\Gamma_{12}+\Gamma_{34})+2\Gamma_{42}\wedge \Gamma_{31}=
\beta e^1\wedge e^2 + \alpha e^3\wedge e^4
\end{equation}
\noindent where

\begin{equation}
\alpha = \left [ 1/2 \left ( \frac{F}{N} \right ) ^{\dot {}} +
   il\left ( \frac{F}{N^2} \right ) \right ]^{ \dot {}}
\end{equation}

\begin{equation}
\beta = \frac{F}{N^3} \left ( il -1/2\dot{N}\right ) ^2 -\frac{\varepsilon}{N}
+2il\sqrt{\frac{F}{N^3}}\left [ \left ({\sqrt{\frac{F}{N}}} \right ) ^{\dot{}} +il\sqrt{\frac{F}{N^3}} \right ]
\end{equation}

\begin{equation}
\gamma =
1/2\frac{F^{1/2}}{N}[ \frac{2}{N^{1/2}} ( il-1/2\dot{N} )  ( ( {\sqrt{\frac{F}{N}}}) ^{\dot{}}+il\sqrt{\frac{F}{N^3}} ) -
\frac{F^{1/2}}{N}  ( il\frac{\dot{N}}{N}+\ddot{N}+1/2\frac{\dot{N}^2}{N} )  ]
\end{equation}

\begin{equation}
 \delta=- \frac{F}{8N\dot{N}} \left ( \frac{4l^2+\dot{N}^2}{N} \right ) ^{\dot{}}
\end{equation}

\noindent we can obtain the Weyl coefficients $C^{(a)}$,
the traceless Ricci tensor
$C_{ab}$ and the scalar curvature $R$
\begin{equation}
 C^{(3)} = \frac{1}{3}(\alpha + \beta + 2\gamma)
\end{equation}

\begin{equation}
R = 2(\alpha + \beta - 4\gamma)
\end{equation}

\begin{equation}
 C_{12} = \frac{1}{2}(\beta-\alpha)= -C_{34}
\end{equation}

\begin{equation}
 C_{33} = -2\delta = C_{44}.
\end{equation}
\noindent Other curvature components are equal to zero.

We shall consider rigidly rotating perfect fluid in the comoving frame
such that the four-velocity in tetrad and coordinate components reads

\begin{equation}
 u_{1} = 0 = u_{2} ;\hspace{1cm} u_{3} = - u_{4} = 1/{\sqrt{2}}
 \end{equation}

 \begin{equation}
 u^{\mu} =  \sqrt{ \frac{N}{F}} \hspace{0.2cm} \delta^{\mu}_{\tau}
\end{equation}

Then the Einstein equations coupled with B-I theories invariant under
duality rotations and perfect fluid become (Salazar et al., 1987)

\begin{equation}
 R = -8b^2 f(X)+ 8b^2Xf^{\bigtriangledown}(X)+3p-\epsilon
\end{equation}

 \begin{equation}
 C_{12} = -2b^2Xf^{\bigtriangledown}(X) -1/4(p+\epsilon)
 \end{equation}

 \begin{equation}
 C_{33} = -1/2(p+\epsilon)
\end{equation}
\noindent denoting the derivative of $f(X)$ with respect to $X$ by superscript
$\bigtriangledown$.

From the definition

\begin{equation}
 J^{\mu} = \rho{u^{\mu}}
 \end{equation}
\noindent we can infer that $X$ depends only on $r$
\begin{equation}
X=X(r)
\end{equation}

\l and the Born-Infield equations read

\begin{equation}
 [N(B+iD)]^{\dot{}} -2ilf^{\bigtriangledown}(X)(B+iD) = 4\pi iNJ^{\tau}
\end{equation}

Solving Eqs. (31)-(36) we arrive to the equations for the pressure $p$
and the energy density $\epsilon$
\begin{equation}
 p =
2b^2f(X)-\frac{\varepsilon}{N}+\frac{\dot{F}\dot{N}}{2N^2}-
\frac{F}{4N^3}(4l^2+\dot{N}^2)
\end{equation}

\begin{equation}
 \epsilon
=-2b^2f(X)+\frac{\varepsilon}{N}-\frac{\dot{F}\dot{N}}{2N^2}+
\frac{3F}{4N^3}(4l^2+\dot{N}^2)-\frac{F\ddot{N}}{N^2}
\end{equation}

\begin{equation}
 p+\epsilon =-\frac{F}{2N\dot{N}} \left ( \frac{4l^2+\dot{N}^2}{N} \right ) ^{\dot{}} \neq
0
\end{equation}
\l and for the electromagnetic field
\begin{equation}
 B =\frac{a_0 -2lA}{N}b\hskip 1cm  D =\frac{\dot{A}b}{f^{\bigtriangledown}(X)}\hskip 1cm A
= A(r)
\end{equation}
\[
a_0 ,b = constants
\]
\begin{equation}
 f^{\bigtriangledown}(X) = \dot{A}/\sqrt{2X- \left ( \frac{a_0 -2lA}{N}\right ) ^2}\hskip 1cm X =
X(A)
\end{equation}
\l the remaining Einstein equation is
 \begin{equation}
 -2b^2Xf^{\bigtriangledown}(X) =
\left [ -2\varepsilon -\ddot{F}+\frac{2\dot{F}\dot{N}}{N}-
\frac{F(4l^2+\dot{N}^2)}{N^2} \right ] /4N
\end{equation}

For every set of functions $A$, $F$, $N$ satisfying Eq. (42)
 we have  an analytic
solution of the Einstein-Born-Infeld equations coupled with a perfect
fluid. Energy density and pressure are given by (37), (38) and the
electromagnetic field is given by (40).

\v1

\l {\sf III. SOME EXPLICIT EXAMPLES:}

\v1

\l {\sf A. Solution without electromagnetic field}

\v1
When we switch off the electromagnetic field $X=0$,
$A=a_0 /2l$ we recover from Eq. (42) the equation given
by Kramer et al. (1987)
\begin{equation}
-2\varepsilon - \ddot{F} + 2\frac{\dot{F}\dot{N}}{N} - F\frac{(4l^{2} +
\dot{N}^{2})}{N^{2}} = 0
\end{equation}

\v1
\l {\sf B.  Linear electromagnetic field}

\v1
When
\begin{equation}
 f(X) = X
\end{equation}
\l Eq.(42) reduces to
\begin{equation}
(\dot A)^{2} + \frac{(a_{0} - 2lA)^{2}}{N^{2}}
 +\frac{1}{4N}\left [ -2\varepsilon -\ddot{F}
+2\frac{\dot{F}\dot{N}}{N} - F\frac{(4l^{2}+\dot{N}^{2})}{N^{2}} \right ] =0
\end{equation}
\l this solution was reported by Garc\'{\i}a and Tellez (1992).

\v1
\l {\sf C. Linear electromagnetic field coupled with dust}

\v1
When we couple to the gravitational and linear electromagnetic field
dust, characterized by the condition $p=0$, then
\begin{equation}
F = \varepsilon r^{2}+C_{1}r + C_{2}\hspace{2cm} C_1 ,C_2 =constants
\end{equation}
\l and Eq. (42) reduces to
\begin{equation}
\dot{A}^{2} + \frac{(a_{0}-2lA)^{2}}{N^{2}}-\frac{1}{2N}\dot{F}\dot{N}
+\frac{F}{4N^{3}}(4l^{2}+\dot{N}^{2}) + \frac{\varepsilon}{N} = 0
\end{equation}
this solution is not reported in the literature.

\v1
\l {\sf D.  Born-Infeld original theory}

\v1
For the original B-I theory described by
$ f(X) = \sqrt{(1 + 2X)} - 1 $ , Eq.
 (42) reduces to
\begin{eqnarray}
b^{2}\frac{(\dot{A})^{2}+(a_{0}-2lA)^{2}/N^{2}}
{\sqrt{(1-(\dot{A})^{2})[1+(a_{0}-2lA)^{2}/N^{2}]}}+ \\ \nonumber
\frac{1}{4N}\left [ -2\varepsilon -\ddot{F}+\frac{2\dot{F}\dot{N}}{N}-
\frac{F(4l^{2}+(\dot{N})^{2})}{N^{2}}\right ] = 0
\end{eqnarray}
this branch of solutions has been
 not reported in the literature, except for
 $A = cte \neq \frac{a_{0}}{2l}$  (Bret\'on, 1989).

\v1
\l {\sf E. Born-Infeld theory with perfect fluid and given equation of state}

\v1
From Eqs. (37) and (38) we arrive to the general equation
\begin{equation}
(3p + \epsilon)^{\dot{}}=
4b^{2}f^{\bigtriangledown}(X)\frac{{(X
N^{2})}^{\dot{}}}{N^{2}}-\frac{F}{N^{2}}\left [ \ddot{N}-\frac{4l^{2}+
(\dot{N})^{2}}{N}\right ] ^{\cdot}
\end{equation}
with the ans\"{a}tz
\begin{equation}
X = \frac{\varepsilon_{0}^{2} + g_{0}^{2}}{N^{2}}\hspace{2cm} \varepsilon _0,
g_0=constants
\end{equation}
\l and the equation of state
\begin{equation}
3p+\epsilon = cte =\gamma_{0}
\end{equation}
Eq. (49) reduces to
\begin{equation}
\left [ \ddot{N}-\frac{4l^{2}+\dot{N}^{2}}{N} \right ] ^{\dot{}}=0
\end{equation}
\l with general solution
\begin{equation}
 N = \beta_{0}
+\sqrt{\frac{4l^{2}}{\beta_{1}}+
(\beta_{0})^{2}}\, \cosh \sqrt{{\beta_{1}}}(r-\beta_{2})
\end{equation}
\[
\beta _0 , \beta _1 , \beta _2  = constants
\]
\l then from the equation of state we obtain for $F$
\begin{equation}
F=
\dot{N}[F_{0}+\int^{r}\frac{N}{(\dot{N})^{2}}[(\gamma_{0}-4b^{2}f(N^{-2}))N+2\varepsilon ]dy]
\end{equation}
\l on the other hand, from
 \begin{equation}
 X = \frac{\varepsilon _0 ^{2} + g_{0}^{2}}{N^{2}}
 \end{equation}
\l  we have two possibilities for the electromagnetic field $B +iD$:

\l i) Magnetic solution

 $$A =constant \neq \frac{a_{0}}{2l}$$
$$ B = \frac{g_{0}}{N}$$
  $$ D = 0$$
\begin{equation}
4\pi J^{\tau}= \frac{2lg_{0}}{N^{2}}f^{\bigtriangledown}
(\frac{g_{0}^{2}}{N^{2}})
\end{equation}
ii) Generalized NUT solution
$$ {\rm for} \hspace{2cm} A \neq cte$$
$$B+iD = \frac{2b^{2}(\varepsilon _0 ^{2}+g_{0}^{2})}{N}e^{i\phi}$$
$$\phi=2l\int^{r}\frac{1}{N}f^{\bigtriangledown}(\frac{1}{N^{2}})dy$$
\begin{equation}
J^{\tau} = 0
\end{equation}
Case ii) generalizes the NUT type D solution of the Einstein-Born-Infeld
equations (Salazar et al, 1987) when we add a rigidly rotating perfect fluid.

\v1
\l {\sf IV.  CONCLUSIONS}

\v1
We have gotten a wide set of solutions of the Einstein equations coupled
with rigidly rotating perfect fluid and linear and non-linear
electromagnetic field.

\l One of them, given by Eq. (57), generalizes the NUT solution in the
case of Born-Infeld and perfect fluid sources satisfying the equation
of state $3p + \epsilon = constant.$
\vspace{2cm}

\l {\sf REFERENCES}

\v1

\l Bret\'on B. N. (1989). Journal of Mathematical Physics, {\bf 30}, 2607.

\l Carter B. (1968).  Commun. Math. Phys. {\bf 10}, 280. 

\l Garc\'{\i}a, D.A. and  Tellez, J. (1992). Journal of Mathematical Physics, ~{\bf 33},
2254.

\l Kramer et al. (1980). {\it Exact Solutions of
Einstein's Field Equations,}
 Cambridge University Press. 

\l Salazar, H. I., Garc\'{\i}a, D.A.,
and Plebanski, J.F. (1987). Journal of Mathematical Physics {\bf 28}, 2171.

\end{document}